\documentclass[12pt,preprint]{aastex}

\received{28 March 2003}
\revised{}
\accepted{}

\cpright{AAS}{2003}

\shortauthors{Hjorth et al.}
\shorttitle{The DLA towards GRB 020124 at $z = 3.20$}

\slugcomment{ApJ, in press}

\newcommand{\hst}{{\sl HST\/}}
\newcommand{\hete}{{\sl HETE-II\/}}

\newcommand{\nh}{N(\ion{H}{1})} 
\newcommand{\ebv}{{\rm E(B$-$V)}}
\newcommand{\ebvm}{\mathrm{E(B-V)}}
\newcommand{\av}{A_V}
\begin{document}

\title{
Very high column density and small reddening towards GRB~020124 at z = 3.20%
%Damped Ly$\alpha$ Absorption along the line of sight to GRB~020124
%at z = 3.198: reconciling small reddening with a large column density%
%Reconciling modest reddening with a large column density towards
%GRB 020124 at z=3.198
%\title{The reddening-to-gas ratio towards a GRB-selected 
%\mbox{Damped Lyman$\alpha$ Absorber} at z = 3.20: 
%low metallicity or dust destruction towards GRB 020124?%
\footnote{\rm Based on observations 
%\title{The reddening-to-gas ratio towards a GRB-selected 
%\mbox{Damped Lyman$\alpha$ Absorber} at z = 3.20: 
%low metallicity or dust destruction towards GRB 020124?%
%\title{A collimated fireball model for the second localized HETE-2 burst:
%the optically dim/dark burst
%\objectname{GRB~020124}\footnote{\rm Based on observations 
%\title{A collimated fireball model for the optically dim/dark burst
%       \objectname{GRB~020124}\footnote{\rm Based on observations 
                 with the Nordic
                 Optical Telescope, which is operated on the island of
                 La Palma jointly by Denmark, Finland, Iceland,
                 Norway, and Sweden, at the Spanish Observatorio del
                 Roque de los Muchachos of the Instituto de
                 Astrof\'{\i}sica de Canarias, and on observations 
                 collected by the Gamma-Ray Burst Collaboration at ESO 
                 (GRACE) at the European Southern Observatory,
		 Paranal, Chile (ESO Large Programme 165.H--0464)}}

\author{
J.~Hjorth\altaffilmark{2},                     % jens@astro.ku.dk
P.~M\o ller\altaffilmark{3},                   % pmoller@eso.org
J.~Gorosabel\altaffilmark{4,5,6},              % gorosabel@stsci.edu
J.~P.~U.~Fynbo\altaffilmark{7,2},              % jfynbo@phys.au.dk
S.~Toft\altaffilmark{2},                       % toft@astro.ku.dk
A.~O.~Jaunsen\altaffilmark{8},                 % ajaunsen@eso.org
A.~A.~Kaas\altaffilmark{9},                    % akaas@not.iac.es
T.~Pursimo\altaffilmark{9},                    % tpursimo@not.iac.es
K.~Torii\altaffilmark{10},                     % torii@crab.riken.go.jp
T.~Kato\altaffilmark{11},                      % tkato@kusastro.kyoto-u.ac.jp
H.~Yamaoka\altaffilmark{12},                   % yamaoka@rc.kyushu-u.ac.jp
A.~Yoshida\altaffilmark{13,10},                % ayoshida@crab.riken.go.jp
B.~Thomsen\altaffilmark{7},                    % bt@phys.au.dk
M.~I.~Andersen\altaffilmark{14},               % mandersen@aip.de
I.~Burud\altaffilmark{6},                      % burud@stsci.edu
J.~M.~Castro~Cer\'on\altaffilmark{15},         % josemari@stsci.edu
A.~J.~Castro-Tirado\altaffilmark{5}            % ajct@iaa.es
A.~S.~Fruchter\altaffilmark{6},                % fruchter@stsci.edu
%J.~Greiner\altaffilmark{17},                  % jcg@mpe.mpg.de
L.~Kaper\altaffilmark{16},                     % lexk@science.uva.nl
%S.~Klose\altaffilmark{20},                    % klose@tls-tautenburg.de
C.~Kouveliotou\altaffilmark{17},               % Chryssa.Kouveliotou-1@nasa.gov
N.~Masetti\altaffilmark{18},                   % masetti@bo.iasf.cnr.it
E.~Palazzi\altaffilmark{18},                   % palazzi@bo.iasf.cnr.it
H.~Pedersen\altaffilmark{2},                   % holger@astro.ku.dk
E.~Pian\altaffilmark{19},                      % pian@ts.astro.it
J.~Rhoads\altaffilmark{6},                     % rhoads@stsci.edu
E.~Rol\altaffilmark{16},                       % evert@science.uva.nl
N.~R.~Tanvir\altaffilmark{20},                 % nrt@star.herts.ac.uk
P.~M.~Vreeswijk\altaffilmark{8},               % pvreeswi@eso.org
R.~A.~M.~J.~Wijers\altaffilmark{16},           % rwijers@science.uva.nl
E.~P.~J.~van~den~Heuvel\altaffilmark{16}       % edvdh@science.uva.nl
}

\altaffiltext{2}{Astronomical Observatory, University of Copenhagen,
Juliane Maries Vej 30, DK--2100 Copenhagen~\O, Denmark; 
\email{jens@astro.ku.dk}}
\altaffiltext{3}{European Southern Observatory, Karl-Schwarzschild-Strasse 2,
D--85748 Garching bei M{\"u}nchen, Germany}
\altaffiltext{4}{Danish Space Space Research Institute, Juliane Maries Vej 30, 
DK--2100 Copenhagen~\O, Denmark}
\altaffiltext{5}{Instituto de Astrof\'{\i}sica de Andaluc\'{\i}a (IAA-CSIC), 
Apartado 3.004, E--18080 Granada, Spain}
\altaffiltext{6}{Space Telescope Science Institute, 3700 San Martin Drive, 
Baltimore, MD 21218}
\altaffiltext{7}{Department of Physics and Astronomy, University of Aarhus,
DK--8000 {\AA}rhus C, Denmark}
\altaffiltext{8}{European Southern Observatory, Casilla 19001, Santiago 19,
Chile}
\altaffiltext{9}{Nordic Optical Telescope, Apartado 474, 
E--38700 St.~Cruz de La Palma, Canary Islands, Spain}
\altaffiltext{10}{ Cosmic Radiation Laboratory, RIKEN, 2--1, Hirosawa, 
Wako 351--0198, Japan}
\altaffiltext{11}{Department of Astronomy, Kyoto University, Sakyo-ku, 
Kyoto 606--8502, Japan}
\altaffiltext{12}{Department of Physics, Faculty of Science, Kyushu 
University, Chuo-ku, Fukuoka 810--8560, Japan}
\altaffiltext{13}{Department of Physics, Aoyama Gakuin University, 6--16--1, 
Chitosedai, Setagaya 157--8572, Japan}
\altaffiltext{14}{Astrophysikalisches Institut Potsdam, 
D--14482 Potsdam, Germany}
\altaffiltext{15}{Real Instituto y Observatorio de la Armada, Secci\'{o}n de 
Astronom\'{\i}a, 11.110 San Fernando-Naval (C\'adiz), Spain}
%\altaffiltext{16}{Laboratorio de Astrof\'{\i}sica Espacial y F\'{\i}sica 
%Fundamental (LAEFF--INTA), Apartado 50.727, E--28.080 Madrid, Spain}
%\altaffiltext{18}{Max-Planck-Institut f\"ur Extraterrestrische Physik, 
%D--85741 Garching, Germany}
\altaffiltext{16}{Astronomical Institute `Anton Pannekoek', 
NL--1098 SJ Amsterdam, The Netherlands}
%\altaffiltext{20}{Th\"uringer Landessternwarte Tautenburg, 
%D--07778 Tautenburg, Germany}
\altaffiltext{17}{Universities Research Association, Marshall Space Flight 
Center (NASA), Huntsville, AL 35812}
\altaffiltext{18}{IASF/CNR, Sezione di Bologna, Via Gobetti 101, 
I--40129 Bologna, Italy}
\altaffiltext{19}{INAF, Osservatorio Astronomico di Trieste, via Tiepolo,
I--34131 Trieste, Italy}
\altaffiltext{20}{Department of Physical Sciences, University of Hertfordshire, 
College Lane, Hatfield, Hertfordshire AL10 9AB, UK}

\begin{abstract}

We present optical and near-infrared observations of the dim afterglow of 
\objectname{GRB 020124}, obtained between 2 and 68 hours after the gamma-ray 
burst. The burst occurred in a very faint ($R\gtrsim29.5$) Damped Ly$\alpha$ 
Absorber (DLA) at a redshift of $z=3.198\pm 0.004$. The derived column density 
of neutral hydrogen is $\log(N_{H I})=21.7\pm0.2$ and the rest-frame reddening 
is constrained to be $\ebvm < 0.065$, i.e., $A_V < 0.20$ for standard 
extinction laws with $R_V \approx 3$. The resulting dust-to-gas ratio is less 
than $11 \%$ of that found in the Milky Way, but consistent with the SMC and
high-redshift QSO DLAs, indicating a low metallicity and/or a low 
dust-to-metals ratio in the burst environment. A grey extinction law 
(large $R_V$), produced through preferential destruction of small dust grains 
by the GRB, could increase the derived $A_V$ and dust-to-gas ratio. The 
dimness of the afterglow is however fully accounted for by the high redshift: 
If \objectname{GRB 020124} had been at $z=1$ it would have been approximately 
1.8 mag brighter---in the range of typical bright afterglows. 
\end{abstract}


\section{Introduction}
Spectroscopy of the optical afterglows of cosmological GRBs allows detailed 
studies of the chemical and kinematical properties of gas along the lines 
of sight. Independently of the brightness of the host galaxy, the resulting 
insight into the GRB environment and its dust and gas content can provide 
clues to the nature of GRB progenitors as well as the properties of 
high-redshift galaxies. In particular, constraints on the column density of
neutral hydrogen, \nh, can be obtained through Ly$\alpha$ absorption if the 
burst is sufficiently distant to redshift the Ly$\alpha$ line into the 
near-UV/optical domain. Unlike X-ray studies where one infers \nh\
only indirectly based on an assumed metallicity, the derived \nh\ will be 
independent of metallicity. 

In this regard, spectroscopy of GRB optical afterglows bears some
resemblance to that of Damped Ly$\alpha$ Absorbers (DLAs, \cite{wolfe86}),
which are gas-rich absorption systems intervening the lines of sight to 
background sources, such as QSOs. DLAs are chemically enriched and are 
therefore most 
likely caused by gas in, or close to, galaxies (e.g., \cite{lu96} and 
references therein; \cite{pettini97a}). In some cases galaxy counterparts 
of DLAs have been detected directly 
\citep{moller93,moller98,djorgovski96,fynbo99,moller02}. The important 
and interesting difference between QSO DLAs and GRB absorption systems is 
that the lines of sight through QSO DLAs are absorption cross-section 
selected whereas GRB absorbers (at the GRB redshift) are `GRB progenitor 
site' selected.  Therefore, GRB absorbers are likely to probe sightlines 
through different parts of DLA galaxies than QSO absorbers do, and hence 
contribute to a more complete picture of the properties of high-redshift 
galaxies. 

The dust content of DLAs has been constrained mainly by two methods: 
i) by comparing the colors of background QSOs having intervening DLAs 
with the colors of a control sample of QSOs without intervening DLAs 
\citep{pei91}, and ii) by studying the metal abundance ratios, most notably 
Zn vs.\ Si and iron group elements, and inferring the dust depletion by 
comparison with interstellar clouds in the Galaxy 
\citep{lu96,kulkarni97,pettini97b,ledoux02}. The main conclusion of these 
studies is that DLAs in general are metal poor and that the fraction of the 
metals bound in dust grains is roughly the same as, or somewhat lower than,
that found in the Galaxy.

This paper presents optical/near-IR photometry and optical spectroscopy of 
the afterglow of \objectname{GRB 020124}. \objectname{GRB 020124} (burst 
trigger on Jan 24.44531 2002 UT) was the second \hete\ burst \citep{ricker02} 
to be localized to arcsec precision and is the only burst for which a host 
galaxy has not been detected when searched for with \hst\ 
\citep{berger02,bloom02}. 
We determine the redshift of the burst to be 3.20, which is the fourth highest 
redshift known for a GRB to date, and the third highest 
(after \objectname{GRB 000131} 
at $z=4.50$ \citep{andersen00} and \objectname{GRB 030323} at $z=3.37$ 
\citep{vreeswijk03a}) based on afterglow 
absorption lines rather than host galaxy emission lines. We also show that it 
has a very high column density, qualifying it as a DLA (the highest value of 
\nh\ of known GRB afterglows and higher than almost all QSO DLAs). 
We discuss the 
implications for the dust-to-gas ratio of the GRB surroundings, including the 
effects of metallicity and destruction of dust by the GRB itself. 

We assume a cosmology where $H_0 = 70$ km s$^{-1}$ Mpc$^{-1}$, 
$\Omega_m = 0.3$, and $\Omega_\Lambda = 0.7$. For these parameters, 
a redshift of 3.20 corresponds to a luminosity distance of 27.47 Gpc 
and a distance modulus of 47.19. One arcsecond corresponds to 7.55 
proper kpc, and the lookback time is 11.5 Gyr.

\section{Observations\label{SECTION:observations}}

The observing log and photometric results are reported in Table~1 and
summarized below.

\subsection{Optical imaging}

The position of the burst was observed at RIKEN with the 0.25~m f/3.4 
hyperboloid astrograph equipped with unfiltered CCD camera AP7p. 
%The telescope is a part of the automated response system placed at
%RIKEN (Wako, Saitama, Japan) which is designed for studying early
%phase light curves of optical GRB afterglows. 
%Two kinds of telescopes are used in the
%system. One is the 0.25-m f/3.4 reflector (used in the current work)
%while the other is 0.20-m f/4.0 Newtonian reflector. These telescopes
%are equipped with unfiltered cooled CCD cameras. Each telescope is
%mounted on a German-type equatorial mount. The mounts are controlled
%by a PC and are automatically slewed to GRB positions in response to
%GCN alert messages. In 2002 January, the system was in test observation 
%phase and manual observation was made with the 0.25-m telescope.
The field of view was $50\arcmin\times 50\arcmin$ which covered the entire 
$12\arcmin$ radius \hete\ error circle \citep{ricker02}. The observation 
started at Jan 24.530 UT; 126 frames of 20-s exposure were acquired by 
Jan 24.574 UT. PSF photometry was applied to each of the 126 frames. 
The resultant photometric measurements were combined to yield a 3.0$\sigma$ 
detection at the position of the afterglow reported by \citet{berger02}. 

R-band images of the optical afterglow were obtained with StanCam at the 
2.56~m Nordic Optical Telescope (NOT) between Jan 26.035 and Jan 26.076 
2002 UT. I-band images were obtained at VLT-Melipal (UT3)
between Jan 26.311 and 
26.318 UT. R-band images were obtained with FORS1 on VLT-Melipal 
between Jan 27.278 UT and 
27.292 UT. The latter
images have a pronounced background pattern due to instrument
reflections of the very bright background (the moon was $\sim75$~\% 
illuminated). The variation in the pattern across the field is approximately 
3--4~\%. To flatten each individual image we used SExtractor to create a 
smoothed background image which was subsequently subtracted. The five 
background-subtracted images were then combined.

All images were preprocessed using standard tools and calibrated using 
nearby comparison stars from the field photometry of \citet{henden02}.
The derived magnitudes are listed in Table~1.

\subsection{Near-infrared imaging}

Near-IR observations were obtained with ISAAC on VLT-Antu (UT1). The data 
consist of a total of 30 min in the Ks band and 30 min in the Js band
obtained on Jan 26.2 UT and 30 min in the Ks band obtained on Jan 27.2 UT. 
The reduction was carried out with the ``eclipse'' software package 
\citep{devillard97} and the IRAF ``Experimental Deep Infrared Mosaicing
Software'' xdimsum. Eclipse was used to remove effects of electrical ghosts
from science and calibration frames and to construct flat fields and
bad-pixel maps from a series of twilight-sky flats. Sky subtraction and
combination of the dithered science frames were carried out with xdimsum. 
To account for the rapidly varying background, the sky value in each 
pixel was calculated as the running median of the pixel value in the exposures 
taken immediately before and after a given exposure. To minimize the effects 
on the running median of bright objects in neighboring frames, the subtraction 
was carried out in two iterative steps: First, a cosmic-ray cleaned, 
background-subtracted combined image was constructed from which a mask of all
the objects was created. This mask was then registered to the individual 
frames and taken into account in the following running median background
subtraction. Residual background signatures were subsequently removed by
fitting high-order polynomials to the rows and columns of the individual
(masked) frames. Finally, the individual background-subtracted images were
registered and median combined. Excerpts of the resulting images are shown in 
Fig.~1. 

Photometric calibration to J and Ks magnitudes was performed using standard 
star observations of 9106/S301--D and 9149/S860--D \citep{persson98}
on the nights of observation.
The images were astrometrically calibrated using comparison stars from the 
Guide Star Catalogue II (GSC-II)\footnote{The Guide Star Catalogue II is a 
joint project of the Space Telescope Science Institute and the Osservatorio 
Astronomico di Torino. Space Telescope Science Institute is operated by the 
Association of Universities for Research in Astronomy, for the National 
Aeronautics and Space Administration under contract NAS5-26555. The 
participation of the Osservatorio Astronomico di Torino is supported by the 
Italian Council for Research in Astronomy. Additional support is provided by 
European Southern Observatory, Space Telescope European Coordinating Facility, 
the International GEMINI project and the European Space Agency Astrophysics 
Division.} using the WCStools software package by 
D. Mink\footnote{\url{http://tdc-www.harvard.edu/TDC.html}}.
The absolute astrometry has an rms of $0\farcs43$ in the Ks band. The 
resulting coordinates of the infrared afterglow of \objectname{GRB 020124} 
are R.A.~(J2000.0) = 09$^h$32$^m$50.82$^s$, 
Dec.~(J2000.0) $= -11^{\circ}31\arcmin11\farcs0$, consistent with the 
position of the radio and optical afterglow \citep{berger02}.

\subsection{Spectroscopy}

Spectroscopic observations were obtained on Jan 26.3 with FORS1 on
VLT-Melipal using the 300V+10 grism and the order-separation filter GG375 
in long-slit mode with a 1.0 arcsec wide slit. This configuration 
provides a wavelength coverage 3650 -- 7500 \AA\ at a resolution of 13~\AA. 
The total integration time was 1200 sec, evenly divided between two 
exposures. Standard data reduction procedures were performed using MIDAS.
The two separate exposures allowed us to reliably reject 
cosmic-ray events before co-addition. The spectrophotometric standard 
LTT9491 \citep{hamuy94} was observed using the same configuration.

\section{Results}

\subsection{Spectral energy distribution: spectral index and redshift estimate}

The extracted, flux calibrated spectrum is shown in Fig.~2. Due to
the faintness of the object spectrum and the resulting dominance of
systematic errors from sky subtraction close to bright airglow lines, 
the signal-to-noise fluctuates rapidly in the red part of the spectrum. 
What may look like absorption or emission features redwards of 5500~\AA\ 
should therefore not be trusted. The strong absorption line around 5100~\AA\ 
is clearly significant, and several other absorption features bluewards of 
5500~\AA\ are marginally significant at the 3--4 $\sigma$ level.

One or more of the absorption lines seen in the blue part of Fig.~2 could 
be due to Ly$\alpha$, which would indicate a redshift in excess of 2.3. 
Our first objective was therefore to construct the spectral energy 
distribution (SED)
in the optical to determine its slope, and to look for possible large-scale 
signatures (spectral drops) which might indicate the onset of the Lyman 
Forest and/or the Lyman Valley as expected in an object with redshift 
significantly in excess of 2.3. For this purpose we carefully defined a 
series of intervals along the spectrum where systematic errors from 
night-sky emission-line subtraction did not pose any problems. 
The total number of 
counts was determined in each interval of the afterglow spectrum as well as 
of the spectrum of the spectrophotometric standard. 
Uncertainties were calculated directly via propagation of photon statistics.

The resulting fluxes and corresponding uncertainties (after correction for
airmass) in each bin are plotted in Fig.~3. In Fig.~3a the minimum-$\chi^2$ 
pure power-law fit ($f_{\nu} \propto \nu^{-\beta}$; $\beta = 2.36\pm0.23$; 
reported uncertainties are 1$\sigma$ errors throughout this paper) is
overplotted, but is seen to be a very poor fit to the data points. In 
particular the data point at 5100~\AA\ falls many standard deviations below any 
power-law fit. Such a large drop in that wide a bin can effectively only be 
caused by a damped Ly$\alpha$ line at a redshift of about 3.2. At this 
redshift one expects to see not only the Ly$\alpha$ line but also a 
significant drop due to the Lyman Forest, and the signature of the onset of 
the red slope of the Lyman Valley \citep{moller90}. In Fig.~3b we therefore 
plot the same data points, but here we overlay a model power law including 
the predicted absorption due to the intervening intergalactic medium at 
$z=3.2$. The power law again represents the minimum-$\chi^2$ fit and has an 
index of $\beta = 1.32\pm0.25$. Comparing the $\chi^2$ of the fits in Fig.~3a 
and Fig.~3b, we find that even if we ignore the presence of the Ly$\alpha$ line 
(i.e.\ consider only the large-scale SED), the $z=3.2$ model SED represents 
an acceptable fit while the pure power law is rejected at the 98.6~\% 
confidence level ($\Delta \chi^2=16.4$ for 20 degrees of freedom (dof)).

\subsection{Redshift and \ion{H}{1} column density of the Ly$\alpha$ line}

Following the determination of the underlying power-law index and
approximate redshift, we proceeded to fit the Ly$\alpha$ line itself. 
The part of the spectrum useful for line-fitting is shown again in Fig.~4, 
here normalized by division with the fitted power-law. Overplotted are model 
absorption spectra with 
$z=3.198$. The model spectra include the Lyman series, \ion{Si}{2}, 
\ion{Si}{3}, and \ion{O}{1} lines. On the blue side of the Ly$\alpha$ line we 
see the expected line-blanketing by the Lyman Forest. At this resolution,
lines in the Lyman Forest are therefore not useful for a precise 
redshift determination, and we shall rely only on the 
\ion{Si}{2}\,$\lambda 1260$ and \ion{O}{1}\,$\lambda 1302$ lines at 
5294~\AA\ and 5468~\AA\ for this purpose. The first line is an unblended 
\ion{Si}{2}\,$\lambda 1260$ line providing a best-fit redshift of 
$3.200\pm 0.004$. The \ion{O}{1}\,$\lambda 1302$ line is blended with a much 
weaker \ion{Si}{2}\,$\lambda 1304$ line 
and provides a redshift of $3.196\pm 0.004$. The combined redshift of these 
two lines, including wavelength calibration errors, is $z_{\rm metal} = 
3.198\pm 0.004$ which we shall adopt as the systemic redshift of the absorber.

Because of the line-blanketing on the blue side of the Ly$\alpha$ line, only 
the red part of the line profile could be used to constrain the \ion{H}{1} 
column density of the absorber. Within the uncertainty of the redshift we 
found that acceptable fits could be obtained in the range 
$\log(N_{H I})=21.7\pm0.2$. The estimated $1\sigma$ range is marked in Fig.~4 
as dotted lines.
%, and the model spectra include lines of the Lyman series, 
%\ion{Si}{2}, \ion{Si}{3}, and \ion{O}{1}. 
This column density is among the very highest values observed for QSO DLAs
\citep{storrielombardi00}. The absorption system is unlikely 
to be ``intervening'' (i.e.\ unrelated to the GRB host) because of the very 
small probability of such a high-column-density intervening absorber. More 
importantly, a significantly higher redshift of the GRB is ruled out by the 
lack of even stronger Lyman Forest absorption redward of 5200~\AA\ (Fig.~3).

\subsection{Characteristics of the optical decay\label{SECTION:parameters}}

In addition to the data presented here, we include the data reported by 
\citet{berger02} to constrain the R-band lightcurve. In Fig.~\ref{lightcurve} 
we plot all the R-band data points vs.\ the logarithm of time since the GRB. 
The power-law decay index derived from a weighted fit to all the data is 
$\alpha = 1.64\pm0.03$ ($f_{\nu} \propto (t-t_{\rm GRB})^{-\alpha}$). This 
fit is formally rejected with a reduced $\chi^2$/dof = 26/14.
Excluding the late-time HST data we derive an acceptable fit with a decay 
index of $\alpha = 1.49\pm0.04$ and $\chi^2$/dof = 6/12. Fitting only to 
the data prior to 1 day after the burst we derive a decay index of 
$\alpha = 1.45\pm0.06$ with $\chi^2$/dof = 5/10. These results confirm the 
conclusion of \citet{berger02} that the lightcurve is becoming progressively 
steeper (a `break') (see Fig.~\ref{lightcurve}). 

\subsection{Optical/near-infrared SED}

Assuming a power-law decay index $\alpha = 1.49\pm0.04$ 
we can interpolate the RIJKs-band data taken 
around Jan.~26 UT to a common epoch at Jan 26.20 2002 UT. After dereddening the 
contemporaneous RIJKs-band points for Galactic extinction \citep{schlegel98}, 
the colors are $R-Ks=2.93\pm0.26$ and $J-Ks=1.33\pm0.20$. These colors are 
consistent with the ones expected of a GRB afterglow (see the shaded region 
of the color--color diagram in Fig.~2 of \citet{gorosabel02a}).

We have used the intrinsic spectral index obtained from the VLT spectrum 
($\beta = 1.32 \pm 0.25$, 
see \S~3.1) to obtain a mock B-band point by a power-law extrapolation of the 
R-band measurement. In order to infer information on the extinction ($\av$,
\ebv), the BRIJKs-band SED was fitted with a functional form $f_{\nu} 
\propto \nu^{-\beta} \times 10^{-0.4 A_{\nu}}$, where $A_{\nu}$ is the 
extinction in magnitudes at rest-frame frequency $\nu$. $A_{\nu}$ was
parametrized in terms of $\av$ using the three extinction laws reported by 
\citet{pei92}, i.e., for the Small Magellanic Cloud (SMC), Large Magellanic 
Cloud (LMC) and for the Milky Way (MW). For comparison purposes we also 
considered the unextinguished case (pure power law spectrum given by $f_{\nu} 
\propto \nu^{-\beta}$).

As shown in Fig.~6 and Table~2 (third column) the fits of the three 
extinction laws are consistent with the data. The derived extinctions
are consistent with the no-extinction case (in which case 
$\beta=0.91\pm0.14$). Based 
on these fits we can set an upper limit of $\ebvm < 0.18$. This upper limit 
however corresponds to an unrealistic value of $\beta$. Constraining the 
spectral index to be $\beta > 0.5$ and the extinction law to be that of the 
SMC (which provides the best fit to this and most other afterglow SEDs)
we find that $\ebvm < 0.065$. The $\beta$ values derived for the three 
extinguished cases agree with the pure power-law case. 

\section{Discussion\label{SECTION:discussion}}

\subsection{Jet--wind fireball model}
Within the framework of the afterglow synchrotron model \citep{sari98},
the decay index, $\alpha$, and the spectral index, $\beta$, are related
through the slope of the electron energy distribution, $p$.
Given the low extinction inferred from the SED analysis we find that a jet 
($\nu < \nu_c$, the synchrotron cooling frequency) expanding into a wind 
medium ($n\propto r^{-2}$) provides the best fireball model for the data. 
For $\alpha_1 = 1.49\pm 0.04$, this model requires $p=2.32\pm 0.05$, and
$\beta = 0.66\pm 0.03$, consistent with the observational results with
modest extinction. Expansion into a homogeneous medium ($\nu < \nu_c$) is 
not strictly ruled out but requires an unusually high value of $p$: the 
predicted values are $p=2.99\pm 0.05$ and $\beta = 1.00\pm 0.03$ 
(for $\alpha_1 = 1.49\pm 0.04$), consistent with the observational values 
in the absence of extinction.

\subsection{How robust is the redshift determination?}
The spectrum has low resolution and signal-to-noise ratio, but there are 
five independent pieces of 
evidence for the redshift. First, assuming a pure power-law for the spectrum 
provides a spectral index of $2.36\pm0.23$ which is incompatible with $\beta = 
0.91\pm0.14$ determined from the photometry. Second, the strong absorption 
line at 5100~\AA\ cannot be explained by anything else than Ly$\alpha$ at a 
redshift around 3.2. Third, assuming $z=3.2$ and fitting a model with free 
spectral index but now including the predicted Lyman Forest and Valley 
absorption, one obtains a minimum-$\chi^2$ fit for a spectral index of $1.32
\pm0.25$ which is fully compatible with the photometric determination. Fourth, 
the $z=3.2$ fit leads to an improvement in $\chi^2$ over a single power-law 
alone (for the same degrees of freedom) which rejects the single power-law at 
the 98.6~\% level.  Fifth, there are two identified metal lines, and none of 
the lines predicted from the explicit fit (e.g. Ly$\beta$) are inconsistent 
with the observed spectrum (Fig.~4).

\subsection{Dim or bright?}
The fact that \objectname{GRB 020124} was a fairly dim burst \citep{berger02}, 
but fitted by a normal fireball model without excessive extinction 
reinforces the conclusion of \citet{hjorth02} (\objectname{GRB 980613}), 
\citet{berger02} (\objectname{GRB 020124}), and \citet{fox03} 
(\objectname{GRB 021211})
that there is a population of dim, unextinguished GRB afterglows 
that can account for at least some of the large fraction of (`dark') bursts for 
which no optical afterglow is detected \citep{fynbo01a}. In the case of
\objectname{GRB 020124} this is mainly due to the fairly high redshift
(unlike \objectname{GRB 980613} and \objectname{GRB 021211} which
are at redshifts 1.096 \citep{djorgovski99} and 1.006 
\citep{vreeswijk03b}, respectively):
If an afterglow is redshifted from $z_1$ to $z_2$ it is dimmed by
$\Delta \mu + 2.5(\beta-\alpha-1)\log{\left ( (1+z_2)/(1+z_1) \right )}$.
For a median GRB redshift of $z_1 = 1$ and $z_2=3.198$ the relative dimming 
amounts to 1.82 mag (for $\alpha = 1.49$ and $\beta = 0.91$). 
If the afterglow had been 1.82 mag brighter it would have been a typical
bright afterglow: $R\approx 16.7$ after two hours and $R\approx 20.7$ after 
one day (see e.g.\ Fig.~3 of \citet{gorosabel02b} or Fig.~2 of \citet{fox03}). 

\subsection{\nh\ vs.\ E(B$-$V)}
The Galactic relation between the column density of neutral hydrogen and
the reddening is 
\nh/\ebv\ = $4.93\pm0.28\times10^{21}$ cm$^{2}$ mag$^{-1}$ \citep{diplas94}. 
For the LMC and SMC the corresponding ratios are 
\nh/\ebv\ = $2\pm0.5\times10^{22}$ cm$^{2}$ mag$^{-1}$ and
\nh/\ebv\ = $4.4\pm0.7\times10^{22}$ cm$^{2}$ mag$^{-1}$,
respectively \citep{koornneef82,bouchet85}.  In Table~3 we list this ratio 
for known GRBs for which the \ion{H}{1} column density and \ebv\ have been 
measured or constrained from optical spectroscopy and optical/near-infrared
photometry (the values given are for an SMC extinction law). In addition to 
\objectname{GRB 020124} these are 
\objectname{GRB 000301C} \citep{jensen01} and
\objectname{GRB 000926} \citep{fynbo01b,fynbo01c}.
%, and \objectname{GRB 011211} \citep{burud03,jakobsson03}.
We exclude \objectname{GRB 021004} which has several intervening absorbers 
at different redshifts. It is evident that the values for \nh/\ebv\ are not 
larger than expected for galaxies like the LMC or SMC.
%(even the Galactic value is allowed for \objectname{GRB 011211}). 
Taken at face value these results 
indicate that the GRB surroundings are low in dust, because of low 
metallicity (e.g., for a universal dust-to-metals ratio) and/or because of
a low dust-to-metals ratio (a large fraction of the metals being in the 
gas phase).  This conclusion is fully consistent with what is found in QSO 
DLAs \citep{pettini97b}.

It has been suggested that the intense UV and X-ray flux from a GRB may 
photoionize the surrounding gas and destroy dust grains 
\citep{waxman00,fruchter01,perna02}, resulting in changes in the derived 
\nh, $\av$, \ebv, and their ratios. \citet{galama99} studied the inferred 
absorption of soft X-rays in BeppoSAX GRB afterglows and compared it to the 
optical extinction as inferred from a fireball model fit to the available 
data. They claimed evidence for a high \nh/$\av$ ratio and concluded that dust 
destruction may be the cause. The most convincing case for dust destruction
along these lines was made by \citet{galama03} who found a high value of 
\nh/$\av$ for \objectname{GRB 010222}, based on {\it Chandra X-ray 
Observatory} data. At the same time they interpreted a strong (i.e., non-grey) 
far-ultraviolet (restframe wavelength $\sim 1100$~\AA) component in the 
extinction of \objectname{GRB 010222} as evidence for dust destruction. 

In a comparison of the properties of DLAs and GRB absorbers, \citet{savaglio03} 
claimed that \objectname{GRB 990123} ($z=1.60$), \objectname{GRB 000926} 
($z=2.04$), and \objectname{GRB 010222} ($z=1.48$) (of which only 
\objectname{GRB 000926} has a measured \nh\ \citep{fynbo01b}) have higher 
metal column densities and contain more dust than DLAs at similar redshifts.
They further suggested that the relatively low reddening (restframe wavelength 
$>1200$~\AA) observed towards these GRBs may be due to a grey extinction law, 
produced by dust destruction \citep{perna03}.

\cite{castro03} argue that the chromium-to-zinc ratio towards 
\objectname{GRB 000926} indicates that the host is depleted in dust relative
to local values to a similar degree as QSO DLAs at the same redshift. 
This conclusion is consistent with what we have suggested above for 
\objectname{GRB 020124}, namely that the GRB surroundings have a low 
dust-to-gas ratio, similar to QSO DLAs \citep{pettini97b}.
An additional argument in 
favor of this interpretation is that a significant amount of grey extinction 
would make the afterglow intrinsically even brighter than indicated above 
(\S 4.3). For example, assuming a Galactic value of \nh/$\av$, 
log \nh = 21.7, destruction of 2/3 of the dust producing the extinction
in the rest-frame V band \citep{perna03} and a flat extinction curve 
would dim the afterglow by 1 mag. Any non-grey component would 
significantly increase this value.

We conclude that the high value of the ratio between column density and 
optical extinction first found by \citet{galama99} from X-ray spectroscopy 
remains when \nh\ is estimated from optical spectroscopy in GRB afterglows 
with a DLA host and the extinction is estimated from the observed reddening 
of the optical/NIR afterglow.
The dust destruction interpretation originally proposed by \citet{galama99} 
to account for the X-ray result could conceivably also be applied in this 
case. The observed small reddening would be due to a modification of the 
extinction law due to preferential destruction of small grains by the prompt 
UV and X-ray radiation from the GRB, leading to a grey extinction law. 
However, we have found that the straightforward alternative explanation, 
namely that GRBs occur in less chemically enriched environments, similar to 
those of the SMC and QSO DLAs, is fully consistent with the optical 
observations reported here. 
%This may either indicate a less chemically enriched environment such as the 
%SMC (as also found for other DLAs) or may potentially indicate a modification 
%of the extinction law due to preferential destruction of small grains by the 
%prompt UV and X-ray radiation from the GRB, leading to a grey extinction law. 
Future joint optical and X-ray spectroscopy of $1.6 \lesssim z \lesssim 2.5$ 
GRBs combined with accurate multi-wavelength observations of the afterglow, in 
particular in the optical/NIR regime, can be used to distinguish between these 
two possibilities and at the same time fix the metallicity of the burst 
environment.

\acknowledgments

We thank Jochen Greiner, Sylvio Klose, and Kristian Pedersen for useful 
comments.
JPUF acknowledges support from the Carlsberg Foundation. 
JMCC acknowledges the receipt of a FPI doctoral fellowship
from Spain's Ministerio de Ciencia y Tecnolog\'{\i}a.
The authors acknowledge benefits from collaboration within the EU FP5
Research Training Network ``Gamma-Ray Bursts: An Enigma and a Tool''.
This work was also supported by the Danish Natural Science Research Council 
(SNF).

\newpage
\includegraphics{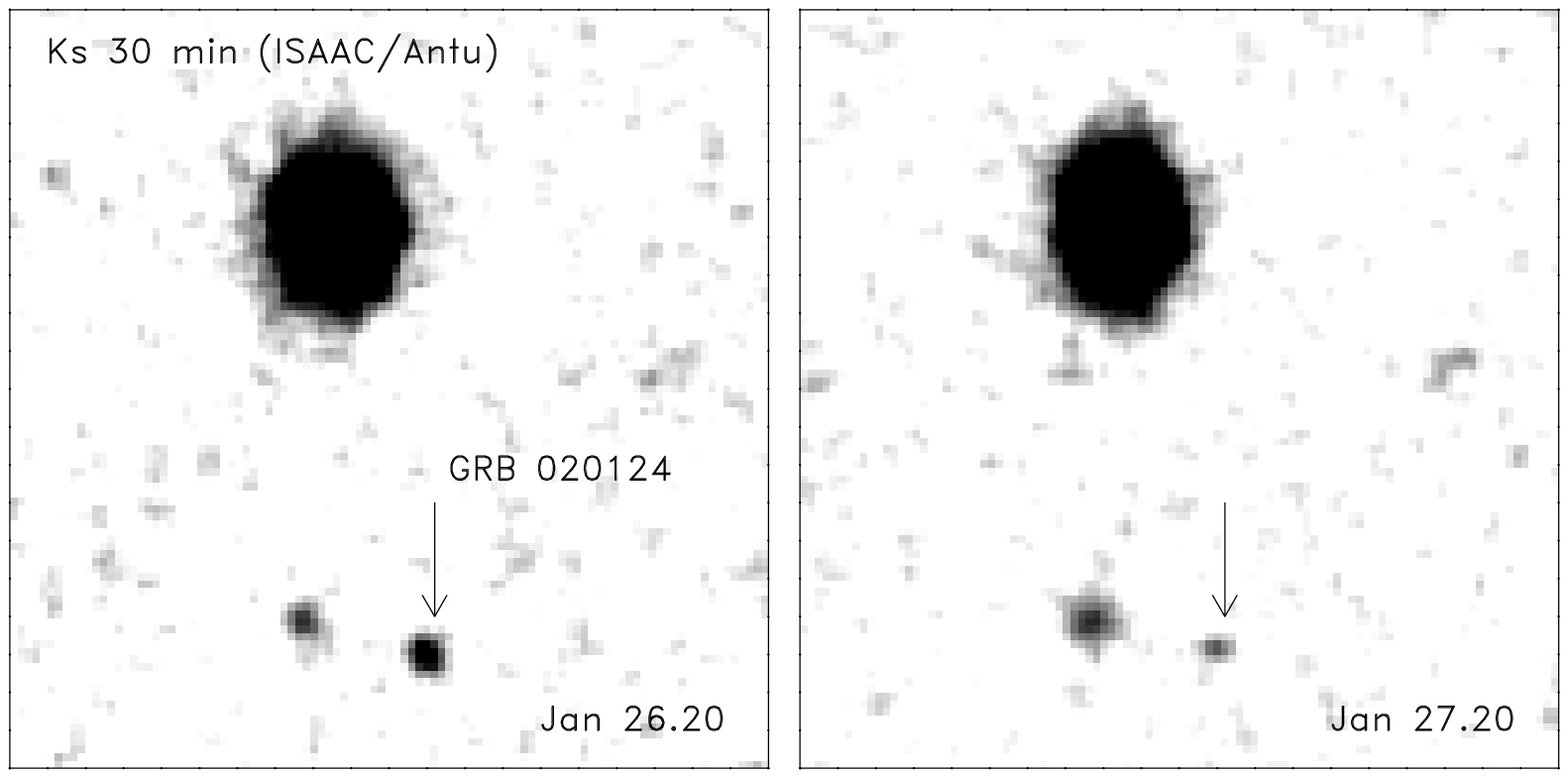}
\figcaption[]{Ks images
obtained with ISAAC on 2002 Jan 26.20 UT (left) and 27.20 UT (right) 
of \protect\objectname{GRB 020124} marked by an arrow. The images are
$15\arcsec \times 15\arcsec$. North is up and east is to the left.
\label{isaac}}

\newpage
\includegraphics[width=15cm]{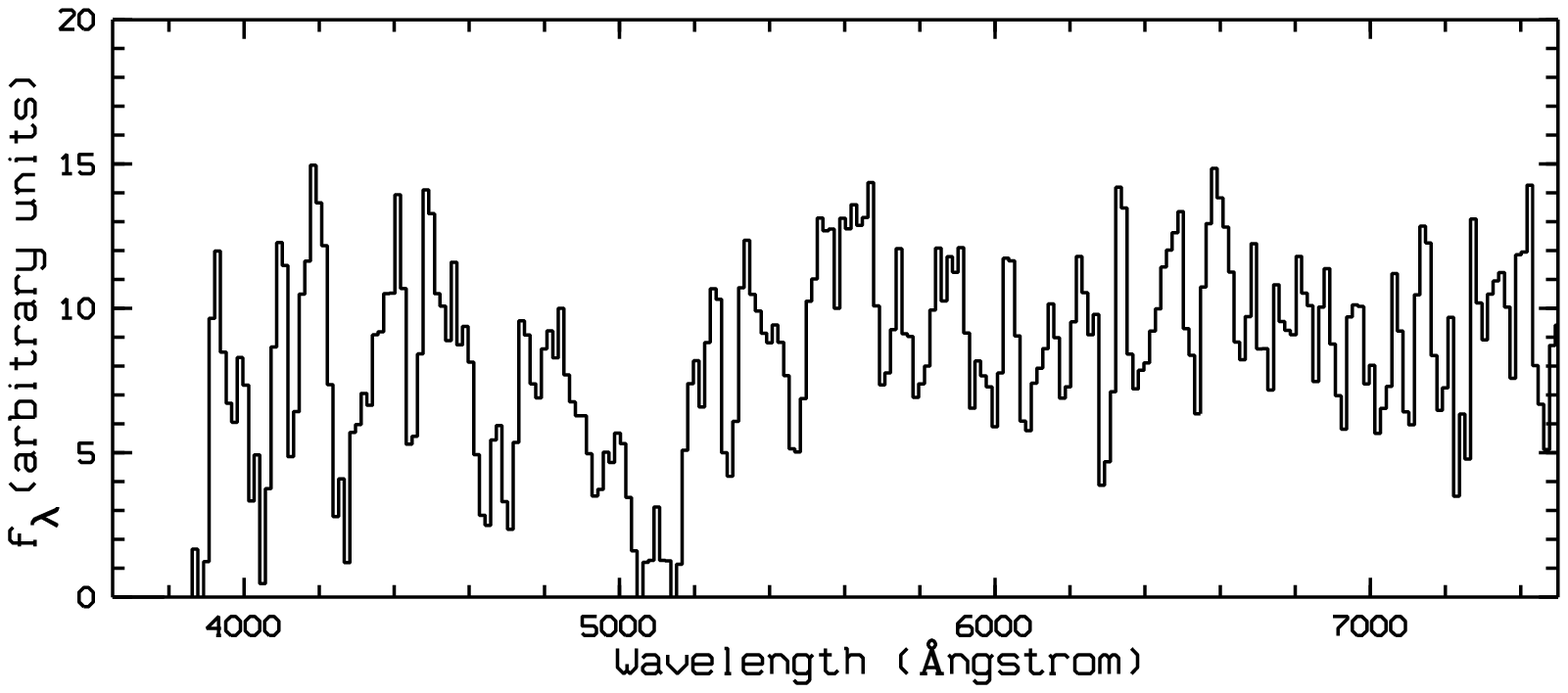}
\figcaption[]{Extracted, flux calibrated FORS1 300V spectrum of 
\protect\objectname{GRB 020124} obtained on 2002 Jan 26.34 UT. 
The ordinate is $f_{\lambda}$ in arbitrary units. From 4000~\AA\ to 
5500~\AA\ the signal-to-noise ratio per resolution element (13~\AA) 
in the continuum grows from 1.8 to 3.5. Longwards of 5500~\AA\ the 
signal-to-noise ratio is completely dominated by sky-subtraction errors 
and therefore varies rapidly from pixel to pixel. In order to obtain 
the best possible value for the spectral slope of the spectrum we have 
identified ``good'' sections of the spectrum as detailed in \S~3.1 and 
shown in Fig.~3.
\label{raw}}

\newpage
\includegraphics[width=15cm]{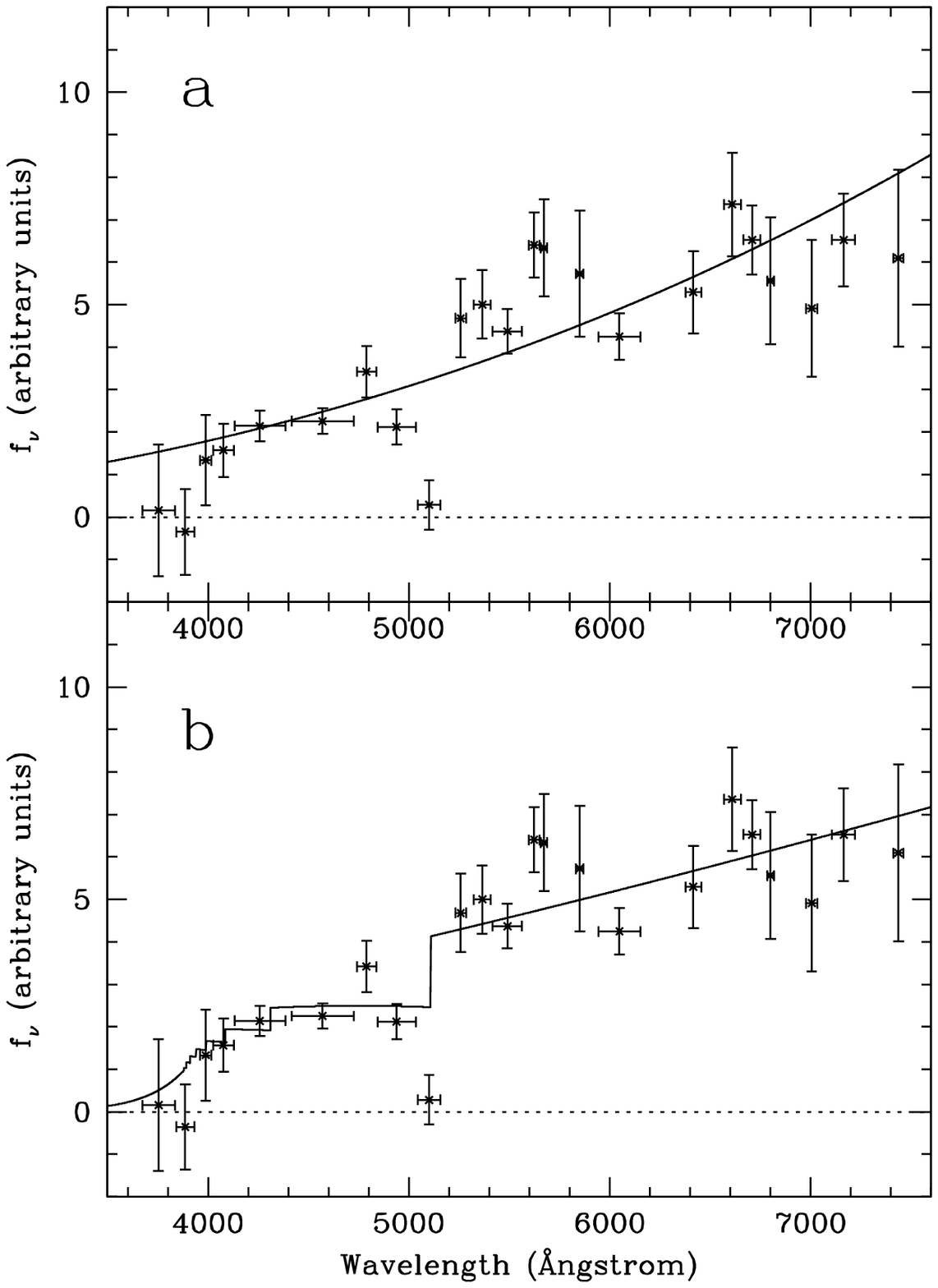}
\figcaption[]{Spectrum of \protect\objectname{GRB 020124} (Fig.~2) summed 
in bins free from strong sky lines. 
(a) A pure power-law does not provide an acceptable fit mostly due to the 
strong absorption line at 5100 \AA. This strong absorption feature can only 
be explained by a damped Ly$\alpha$ line close to $z=3.2$. A power-law fit 
to the data excluding the bin at 5100 \AA\ leads to a spectral slope of
$\beta = 2.36\pm0.23$ incompatible with the optical/near-IR imaging. 
(b) A power-law fit including the effect of the Lyman Forest and Lyman
Valley for a source redshift of $z=3.198$ again omitting the bin at 5100
\AA. The pure power-law fit (a) is rejected at the 98.6~\%
confidence level. The power-law slope of $\beta = 1.32\pm0.25$ resulting 
from the absorbed model fit (b) is consistent with the slope derived from the 
optical/near-IR imaging ($\beta=0.91\pm0.14$). 
\label{spectrum}}

\newpage
\includegraphics[width=15cm]{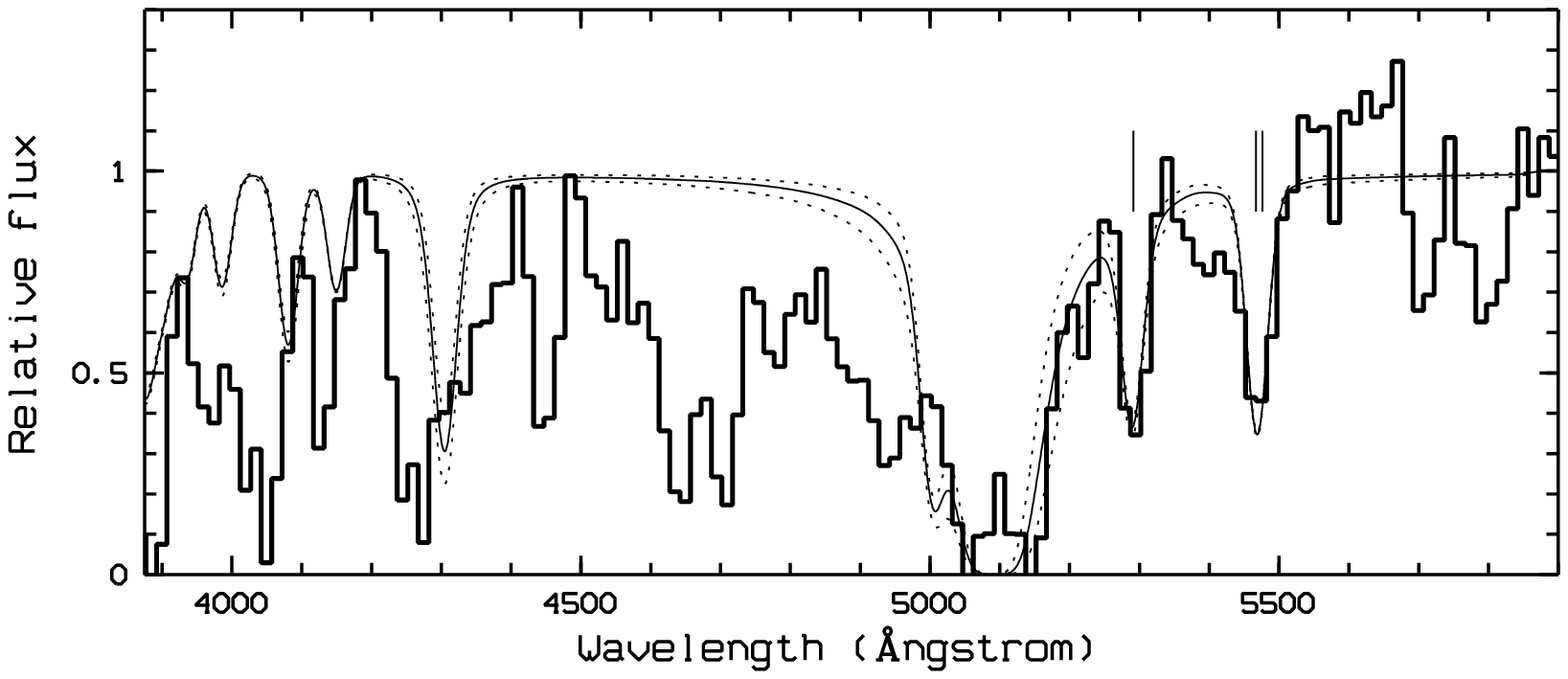}
\figcaption[]{FORS1 300V spectrum and model overlay (solid curve) of a
$z=3.198$ damped Ly$\alpha$ line yielding $\log(N_{H I})=21.7\pm0.2$. 
The estimated $1\sigma$ range is plotted as dotted lines. 
The model has been fit to the red wing of the Ly$\alpha$ line and
the two lines redward of it. These metal lines, identified as
\ion{Si}{2}\,$\lambda 1260$ and \ion{O}{1}\,$\lambda 1302$ 
at 5294~\AA\ and 5468~\AA, were used for the redshift determination. 
For illustrative purposes we
also show the region blueward of the Ly$\alpha$ line. This region
is dominated by Lyman Forest absorption; the data are of
very low signal-to-noise. The predicted model is also shown but
is not fit to the data in this region. The data are consistent 
with the model considering the effect of Lyman Forest absorption. 
%Also indicated are the locations of (left to right)
%Ly$\epsilon$,
%Ly$\delta$,
%\ion{O}{1},
%Ly$\gamma$,
%\ion{O}{1},
%\ion{Si}{2},
%Ly$\beta$,
%\ion{Si}{2},
%\ion{Si}{2},
%\ion{Si}{3},
%Ly$\alpha$,
%\ion{Si}{2},
%\ion{O}{1}, and
%\ion{Si}{2}.
%Bluewards of Ly$\epsilon$ the model drops towards the Lyman limit.
\label{dla}}

\newpage
\includegraphics[width=16cm,clip=]{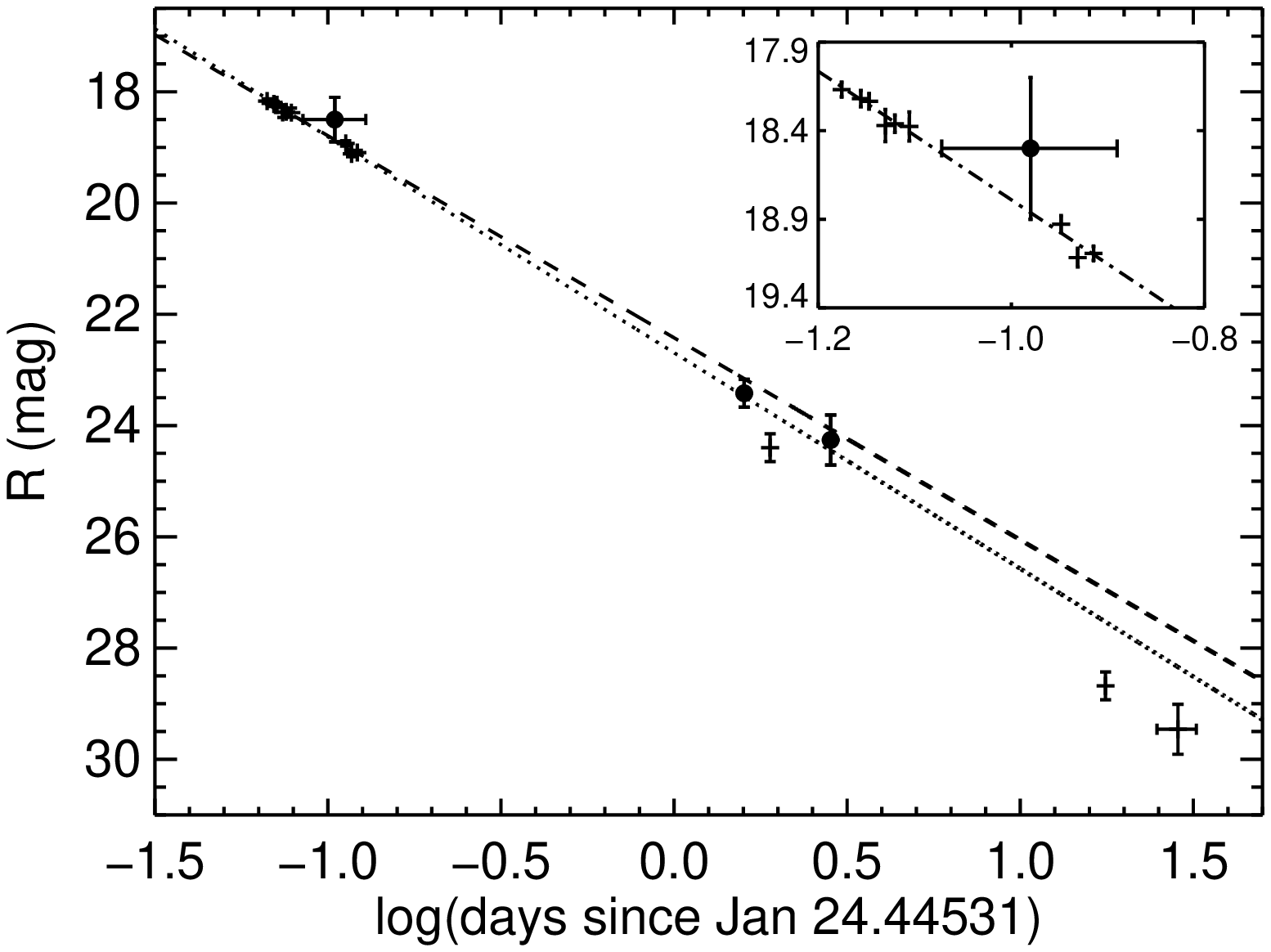}
\figcaption[clip=]{The R-band lightcurve based on our data (filled
circles) and from \citet{berger02}.
The dashed line shows a fit to the early data points only.
The dotted line is the result of a fit including also the two points
from 1 to 3 days after the burst. 
\label{lightcurve}}

\newpage
\includegraphics[width=15cm]{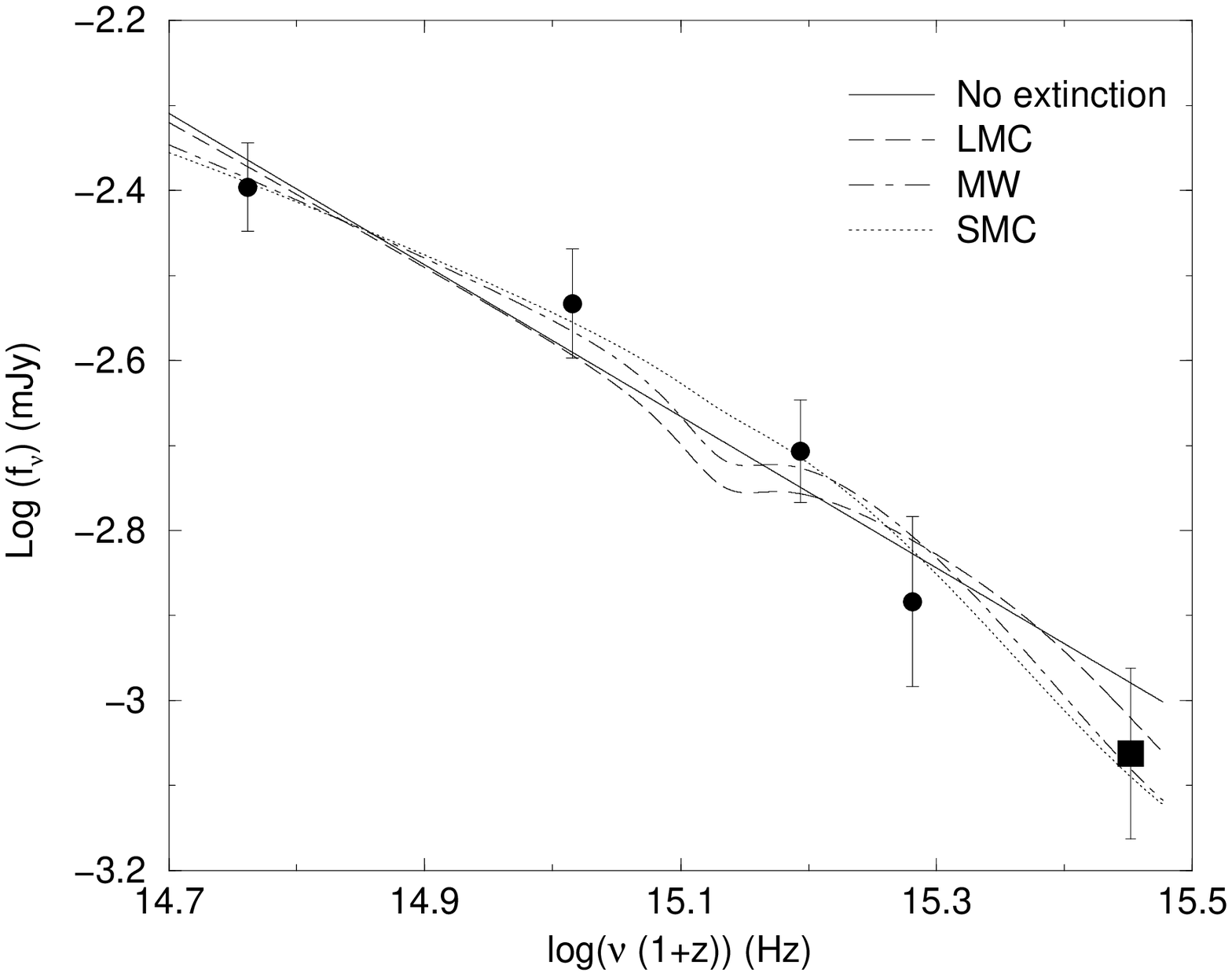}
\figcaption[clip=]{The BRIJKs-band spectral energy distribution of the
afterglow on Jan 26.20 2003 UT. The filled circles represent the 
RIJKs-band measurements. The filled square is the fiducial B-band photometric 
point obtained by extrapolating the R-band point using the intrinsic
spectral slope ($\beta = 1.32\pm0.25$) derived from the VLT spectrum.
The fluxes have been corrected for Galactic extinction \citep{schlegel98}. 
The frequencies are given in the rest frame for $z=3.198$.  
%ZZZ ?
%The extincted SED fits (the three dotted lines) yield very similar
%results to the unextincted case (solid line).
\label{sed}}

\begin{deluxetable}{ccccccc}
%\small
%\footnotesize
%\scriptsize
%\tabletypesize{\scriptsize}
%\tablewidth{7truecm}
%\tablenum{}
\tablecaption{Log of observations and photometry of the afterglow of 
\protect\objectname{GRB 020124}
\label{tbl-1}}

\tablehead{
\colhead{Date} &
\colhead{Filter\tablenotemark{a}/Grism} &
\colhead{Exposure time}&
\colhead{Telescope\tablenotemark{b}}&
\colhead{FWHM\tablenotemark{c}} &
\colhead{Brightness\tablenotemark{d}} \\
\colhead{(Jan 2002 UT)}&
\colhead{}&
\colhead{(sec)} &
\colhead{}&
\colhead{(arcsec)} &
\colhead{(mag)} 
}
\startdata
24.55 & unfiltered& $126\times 20$  & RIKEN   & 2.2  & $18.5^{+0.4}_{-0.3}$   \\
26.04 &        R  & $6\times 600$   & NOT     & 1.14 & $23.42\pm0.25$         \\
26.20 & Ks        & $30\times 60$   & Antu    & 0.50 & $20.55\pm0.13$         \\
26.23 & Js        & $20\times 90$   & Antu    & 0.68 & $21.94\pm0.16$         \\
26.32 &        I  & $2\times60+120$ & Melipal & 0.74 & $22.94\pm0.15$         \\
26.34 & 300V      & $2\times 600$   & Melipal & 1.2  &                        \\
27.20 & Ks        & $30\times 60$   & Antu    & 0.56 & $21.92^{+0.37}_{-0.28}$\\
27.28 &        R  & $5\times 180$   & Melipal & 0.66 & $24.26^{+0.45}_{-0.32}$\\
\enddata

\tablenotetext{a}{The broad-band optical (R, I) filters used were Bessel 
filters \citep{bessel90}.}
\tablenotetext{b}{RIKEN 
represents a 0.25~m f/3.4 hyperboloid astrograph (Wako, Saitama, Japan)
equipped with an unfiltered CCD camera;
NOT is the 2.56~m Nordic Optical Telescope (La Palma, Canary Islands, Spain)
equipped with StanCam; 
Antu and Melipal are the 8.2~m Unit Telescopes 1 and 3 on ESO's VLT at 
Paranal Observatory, Chile, equipped with ISAAC and FORS1, respectively.
}
\tablenotetext{c}{Measured seeing full width at half maximum of point-like
objects in the field.}
\tablenotetext{d}{Uncorrected for Galactic extinction. The brightnesses
are reported as R$_C$, I$_C$, J, Ks magnitudes.}

\end{deluxetable}

\begin{deluxetable}{cccc}
\tablecaption{SED fitting results}
\label{tablesed}
\tablewidth{10truecm}
\tablehead{
\colhead{Extinction law} &
\colhead{$\chi^2/\mathrm{dof}$} &
\colhead{\ebv}&
\colhead{$\beta$}
}
\startdata
 MW            & 1.22 & 0.045$\pm$0.101 & 0.69$\pm$0.50 \\
 LMC           & 0.56 & 0.098$\pm$0.079 & 0.29$\pm$0.52 \\
 SMC           & 0.31 & 0.071$\pm$0.051 & 0.31$\pm$0.43 \\ 
 No extinction & 0.78 & 0               & 0.91$\pm$0.14 \\

\enddata
\end{deluxetable}

\begin{deluxetable}{lccccl}
%\small
%\footnotesize
%\scriptsize
%\tabletypesize{\scriptsize}
%\tabletypesize{\small}
%\tablewidth{7truecm}
%\tablenum{}
\tablecaption{Column density and reddening for GRB absorbers
\label{tbl-2}}
\tablewidth{16.4truecm}
\tablehead{
\colhead{GRB} &
\colhead{$z$} &
\colhead{log N(\ion{H}{1})} &
\colhead{\ebv}&
%\colhead{A$_V$}&
\colhead{N(\ion{H}{1})/\ebv} &
%\colhead{N(\ion{H}{1})/$\av$} &
\colhead{Reference} \\
\colhead{} &
\colhead{} &
%\colhead{} &
%\colhead{} &
\colhead{} &
\colhead{(mag)} &
\colhead{(10$^{21}$ cm$^2$ mag$^{-1}$)} &
\colhead{}
}
\startdata
000301C & 2.040 & $21.2\pm0.5$ & $0.031\pm0.014$ & $51^{+113}_{-42}$  & Jensen et al.~2001\\
000926  & 2.038 & $21.3\pm0.2$ & $0.062\pm0.020$ & $32^{+21}_{-16}$   & Fynbo et al.~2001b,c\\
%011211  & 2.140 & $\sim20.2\pm0.3$   & $<0.054$ & $>1.5$      & Burud et al.~2003; Jakobsson et al.~2003 \\
020124  & 3.198 & $21.7\pm0.2$ & $<0.065$         & $>49$             & this paper  \\
% 98 + 57 - 36
%021004 & 2.3XX & $<20.0$      & $0.084\pm0.013$ & $1.X$& M\o ller et al.~2002; Holland et al.~2003\\
\enddata

\end{deluxetable}

\end{document}